# Choline chloride as a nano-crowder protects HP-36 from urea-induced denaturation: Insights from Solvent Dynamics and Protein-Solvent interaction

Atanu Maity, Soham Sarkar, Ligesh Theeyancheri, and Rajarshi Chakrabarti*

Department of Chemistry, Indian Institute of Technology Bombay, Powai, Mumbai-400076
E-mail: rajarshi@chem.iitb.ac.in

## Abstract

Urea at sufficiently high concentration unfolds the secondary structure of proteins leading to denaturation. In contrast, Choline Chloride (ChCl) and urea, in 1:2 molar ratio form a deep eutectic mixture, a liquid at room temperature and protect proteins from denaturation. In order to get a microscopic picture of this phenomenon, we perform extensive all-atom molecular dynamics simulations on a model protein HP-36. Based on our calculation of Kirkwood-Buff integrals, we analyze the relative accumulation of these osmolytes around the protein. Additional insights are drawn from the analyses of the translational and rotational dynamics of solvent molecules and also from the hydrogen bond auto-correlation functions. In the presence of urea, water shows slow subdiffusive dynamics around the protein backbone as a consequence of stronger interaction of water molecules with the backbone atoms. Urea also shows subdiffuive motion. Addition of choline further slows down the dynamics of urea restricting its inclusion around the protein backbone. Adding to this, choline molecules in the first solvation shell of the protein shows the strongest subdiffusive behavior. In other words, ChCl behaves as a nano-crowder by excluding urea from the protein backbone and thereby slowing down the dynamics of the water layer around the protein. This prevents the protein from denaturation and makes it structurally rigid which in turn is supported by the smaller radius of gyration and root mean square deviation values of HP-36 when ChCl is present in the system.

# 1. Introduction

In some specific organisms, there are some naturally occurring osmolytes which protect the cellular component from different environmental stresses such as high temperature, high salt concentration, desiccation, etc.[1] These are mostly polyols,[2] certain amino acids,[3] and methylamines e.g. trimethylamine N-oxide (TMAO),[4] Triethyl ammonium acetate (TEAA),[5] etc. Due to their stabilizing effect on protein, several other osmolyte molecules like choline chloride,[6] choline-o-sulfate,[7] trehalose [8,9], etc., though not naturally occurring, are used for the preservation of biomolecules for clinical and industrial purposes. Therefore it is important to understand the microscopic reason behind the mechanism of the protecting activity of these osmolytes. For example, how a protecting agent impacts the surrounding aqueous environment of protein so that the structure of the protein remains unaffected or how a protecting agent can influence protein's interaction with denaturant like urea. The effects of osmolytes have been studied in restricting the unfolding of protein from chemical denaturation and denaturation caused by the effect of temperature and pressure are studied by. For example, using small peptides and proteins, Cho *et al.* showed that TMAO can inhibit the folding to unfolding transition.[10, 11] Sarkar *et al.* showed that TEAA and ChCl can restrict the urea induced unfolding of HP-36.[6] Although there are studies that investigated the change in solvent properties in the presence of co-solvent, the detailed understanding of solvation dynamics and its impact on protein's conformation is yet to be explored. The surrounding of protein in cellular environment consists mainly of water molecules along with ions and other important co-factors. The water molecules in the immediate vicinity of protein influence the dynamics and function of proteins.[12, 13] These surrounding water molecules are known ~~as~~ to form a hydration layer. The hydration layer around the protein molecule interacts directly with different parts of the protein molecule. For example, interaction with the solvent-exposed heterogeneous protein surface, with the polar amino acid side chains, and with the backbone atoms. In several cases, hydration of protein is required to impart structural flexibility.[14] Therefore understanding the hydration of protein and its impact on its function demand a prerequisite knowledge of protein-water interactions, which include structural changes and local dynamics of water in the vicinity of a protein backbone and side-chain up to at least couple of hydration layers. Frauenfelder and group suggested that the conformational fluctuations in protein structure are dominated by the dynamics of solvent (primarily water) molecules.[15–18] The large-scale movement of protein follows the fluctuations of bulk water, whereas the small-scale movements are correlated with the fluctuations of the hydration layer. The simplest explanation that can be drawn is that the hydration layer is too rigid to rotate along with the motion of the protein.[19] One could find the coexistence of both slow (from tens of picoseconds to nanoseconds) and fast (in the order of few picoseconds) dynamics of water in the hydration layer. The existence of this bimodal dynamics can be explained in terms of 'bound' (hydrogen-bonded to a macromolecular surface) and 'free' water molecules in the layer.[20] Bhattacharya and group have shown important dynamics of water in different time-scale.[21–23] Though several experimental methods like Nuclear Magnetic resonance (NMR),[14,24] Quasi-elastic neutron scattering (QUENS)[25,26], etc. show promising results in characterizing solvation dynamics, all-atom molecular dynamics (MD) simulation is the most useful method that can reveal the microscopic details of solvation dynamics. MD simulations can successfully trace the key features of protein hydration water near the surface.[27] For example, computational study has confirmed the freezing of the protein movements that results in faster solvation at the W7 site of apo myoglobin. Zhong *et al.* observed the pivotal role of the charged/polar residues as well as the side-chain fluctuations in the slowness.[28] It is rather easy to understand the cause of fast freezing of side-chain motion, which is originated from the removal of slow component from time dependence of solvation energy. Gu and Schoenborn studied the hydration of ribonuclease-A at room temperature and at high extent of hydration.[29] They showed continuous translational and rotational motions in the layer of water. Both rotational and translational diffusion coefficients of water molecules are correlated with the residence time as the translational diffusion is a direct measure of the rigidity of the layer. The rotational relaxation of water molecules in the vicinity of lysozyme was found to be 3-7 times slower than that in the bulk which depends on how the hydration shell was defined surround the protein molecule. Bandyopadhyay *et al.* found a slow component of water near the HP-36 model peptide in the solvation dynamics study.[30] Hydrogen bond (HB) network provides some fascinating features to water dynamics, hence its study can be used as a tool to understand the origin of motions of water molecules both in bulk and near the vicinity of biomolecules. Each water molecule can form four hydrogen bonds. Luzar and Chandler elucidated different parameters of HB lifetime dynamics in neat water, and later this idea has been further extended to explore hydrogen bond dynamics in complex situations like proteins and electrolytes and micelle surfaces.[31,32] In MD simulation, usually the geometric or the energetic criteria are used to define an HB. Classical atomistic MD simulations revealed that the relaxation of water-water HB is much faster than that of the protein-water HB. A correlation is established between protein water HB dynamics with the biological activity. Thus, important insight can be drawn from studies of solvation dynamics which can shed light on the role of different factors influencing the water dynamics to modulate the behavior of protein. To the best of our knowledge, such studies on the dynamic behavior of co-solvents, such as- translational motion, rotational auto-correlation and HB dynamics of water in the presence of co-solvent such as urea and choline are lacking.

In this investigation, aiming at the elucidation of the role of bulky choline in protecting the urea-induced denaturation of protein we consider a model system HP-36. HP-36 is a well-characterized model protein used to study protein folding-unfolding phenomena frequently and it unfolds by the addition of 8M urea in the system. Using this as a model, important insights are drawn regarding the movement of the constituent from the analyses of the solvent dynamics around the protein using the calculation of MSD, rotational auto-correlation, and HB auto-correlation function. Kirkwood-Buff integral (KBI) is also calculated that quantifies the preferential interaction of protein for one solvent/co-solvent over the other. These analyses reveal that in the presence of urea, the dynamics of water slows down. On the other hand, the addition of bulky choline restricts the inclusion of urea to the protein backbone and slows down urea dynamics by acting as a nano-crowder.

The manuscript is arranged as follows. In section 2 we provide the simulation details, in section 3 structural changes of HP-36 is discussed, distribution and dynamics of solvents and co-solvents around the protein (HP-36) are provided in section 4 and 5 respectively, section 6 describes how choline chloride (ChCl) imposes effect of crowding on the protein.

## 2. Methods

### 2.1. Modeling

The structure of the 36-residue villin headpiece is extracted from the deposited NMR structure (PDB Id: 1VII).[33] The C-terminal and N-terminal residues are protected by adding amide and acetyl group respectively to avoid bare charge interaction. HP-36 consists of three helices (Figure1). The protein is then put into a cubic box of volume 245 nm$^3$ and the box was filled with suitable solvent and co-solvent depending on the system chosen for investigation. Four systems are generated considering urea and choline chloride as co-solvents along with water i.e. PW: protein in neat water, PWU: protein in 8M urea and water, PWC: protein in 4M choline chloride and water, PWUC: protein in 2:1 ratio of urea and choline chloride and water. Details information regarding each system is provided in the supplementary information (Table S1).

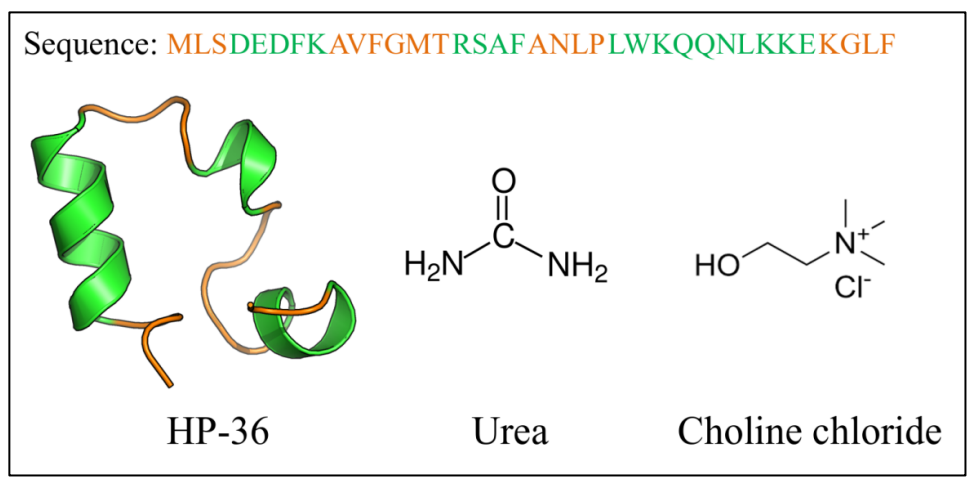

**Figure 1:** Structure of HP-36 (PDB id: 1VII), urea and choline chloride (ChCl). The helices and the coil in the 3D representation of HP-36 are colored in green and orange respectively. The amino acid sequence is also shown and colored accordingly.

## 2.2. Simulation details

OPLS-AA force field is implemented to model the protein (HP-36) and the co-solvent (ChCl and urea).[34] For water, SPC/E model is used.[35] The parameters for small molecules are taken from a previous study.[6] For the removal of initial steric clashes, 5000-steps energy minimization is performed for each system using the steepest descent method.[36] Subsequently, a 200 ps equilibration in NVT ensemble is performed to equilibrate each system at 310K and to avoid any kind of void formation in the box followed by a 5 ns equilibration at isothermal-isobaric (NPT) ensemble to attain a steady pressure of 1 atm. This period of equilibration is found to be sufficient enough for the convergence of properties like pressure, temperature for each system. The temperature was kept constant at 310 K by applying the V–rescale thermostat[37] and pressure is maintained to be at 1 atm using Parrinello-Rahman barostat[38] with a pressure relaxation time of 2ps used for the attainment of desired pressure for all simulations. The production runs for 200 ns with a time step of 2 fs are performed for PW, PWU and PWC systems whereas PWUC system is simulated for 500ns using GROMACS 5.0.5.[39] The longer trajectory length considered for protein in the ternary mixture (W+U+C) is to equilibrate the mixture properly. Short-range Lennard–Jones interactions are calculated using the minimum image convention.[40] For estimating non-bonding interactions including electrostatic as well as van der Waals interactions, a spherical cut-off distance 1 nm is chosen. Periodic boundary conditions have been used in all three directions for removing edge effects. SHAKE algorithm[41] is applied to constrain bonds involving hydrogen atom of the water molecules. Long-range electrostatic interactions are calculated using the particle mesh Ewald (PME) method.[42] The frames in the trajectory are saved at a frequency of 2ps for analysis. To extract different structural and dynamic properties in-built modules of GROMACS 5.0.5 [39] and some in-house scripts are used. For visualization purpose VMD 1.9.3 [43] is used.

# 3. Structural changes of HP-36 in the presence of solvent and co-solvent

## 3.1. Time evolution of α-helices

In this study of urea induced denaturation of protein secondary structure and its subsequent counteraction by the protecting osmolytes, we have used a helix rich model protein[33] with three proper helices in its structure. Hence as an initial sign of the destruction of the protein structure or its preservation, number of residues contain α-helices can be a good representative of the folding-unfolding equilibrium. Figure 2 represents the time evolution of the percentage of α-helicity for different systems considered in this study. HP-36 is found to maintain its native structural fold during the 200ns simulation in water (PW system, presented in black). The percentage of α-helicity for PW system is maintained ~55% throughout the system. In the presence of urea (PWU system, presented in red) there is a gradual loss of the α-helices starting around 60ns and this denaturation process gets completed around 160 ns of the trajectory. This 100ns time duration can be attributed as the transition time region, where the percentage of α-helicity of the model protein drops from ~55% to ~15%. For PWU system, 160ns onwards the protein exists only in the coil-like motif with a percentage of α-helicity tends to 0% as time progresses. While considering other systems, namely PWC (presented in blue) and the ternary mixture PWUC (presented in green), we find that the number of amino acid residues containing the α-helices remains almost constant hence the percentage α-helicity shows a constant value ~55% throughout the trajectory.

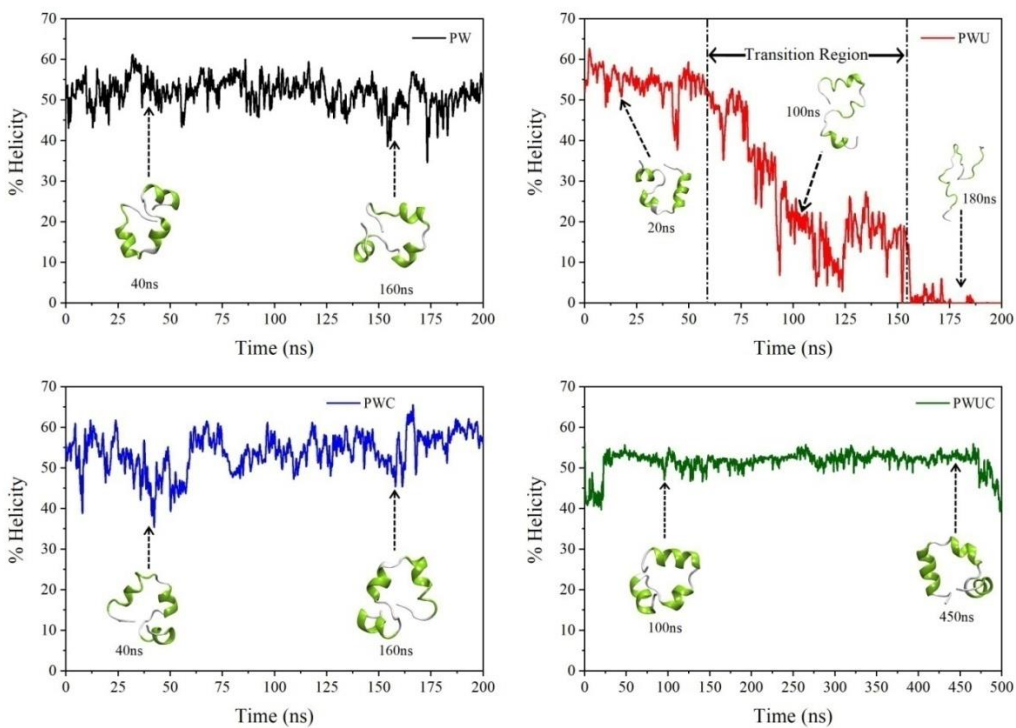

**Figure 2:** Time evolution of the percentage of α-helicity of HP-36 in different systems considered in the study. Relevant snapshots of the protein secondary structure are also provided within the plots. The helices and the coil in the 3D representation of the small protein are colored in green and white respectively.

## 3.2. Time evolution of Secondary structure of the model protein

We analyze the secondary structure of the model protein for four different systems using the 'do-dssp' utility of GROMACS 5.0.5. [39] The results are put in Figure 3(a-d), which highlights the changes in its helical properties, helix-coil transition, and the presence of other secondary structural elements. It is evident from Figure 3b (PWU system) that the considered protein structure undergoes significant helix-coil transitions in the presence of urea during the 200ns simulation. In the plot, helices are represented in magenta color and the coil-like structural motifs are represented in green color. It appears that the process of denaturation starts from the second helix (residue 15-18) and gradually get entirely distorted to a coil structure in the absence of any protecting osmolytes (choline chloride here). Figure 3a, c, and d show expected retention of the initial helical structure of HP-36 in water (PW), water/choline chloride (PWC) and in the ternary mixture (PWUC) throughout the trajectory respectively. The secondary structure under the deep eutectic condition (Figure 3d) shows enhanced stability of the constituent α-helices almost throughout the simulated trajectory.

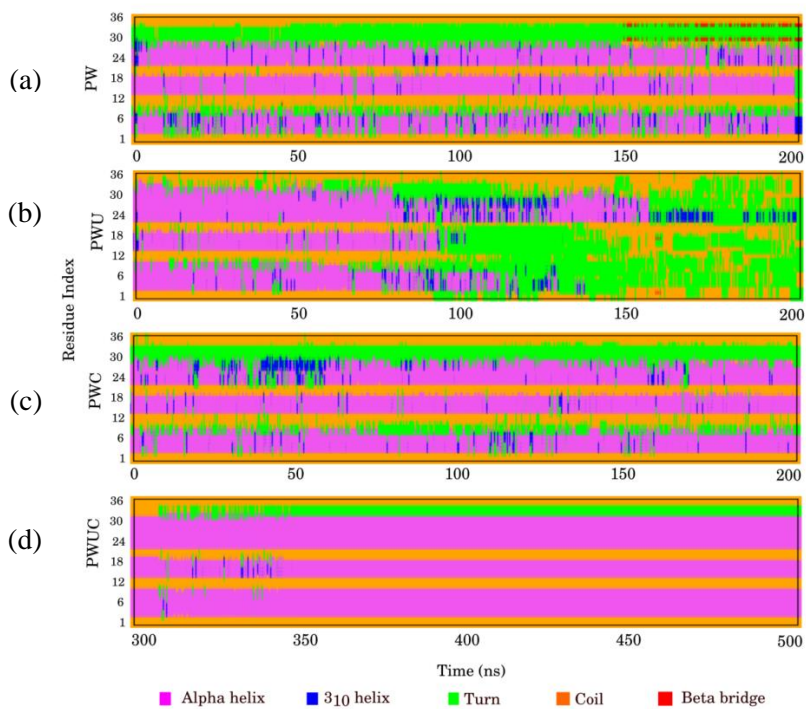

**Figure 3:** Time evolution of the secondary structure elements of HP-36 with different combinations of co-solvents (discussed in detail in the method section) considered in this investigation. For the PWUC system total 500ns trajectory is generated, within which the last 200ns is taken into consideration.

## 4. Distribution of solvent and co-solvent around protein

### 4.1. Calculation of Radial distribution function

Radial distribution function (g(r)) gives an average picture of the distribution of water or osmolytes (choline or urea) around the protein. g(r) of water, urea, and choline is plotted in Figure 4. For constructing the g(r) plots, we considered protein backbone and water/urea/choline as two groups.

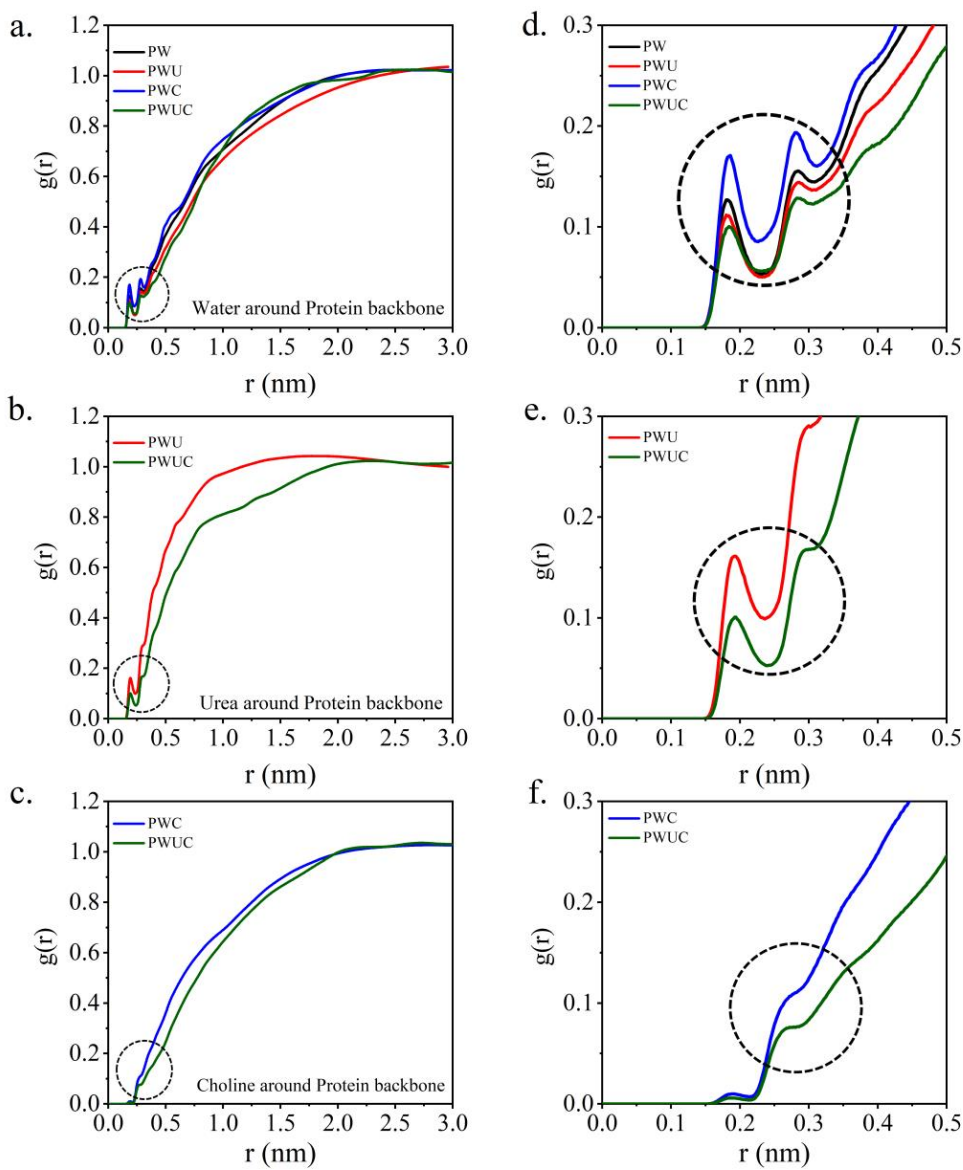

**Figure 4:** Radial distribution functions (g(r)) of a) water, b) urea and c) choline around HP-36 backbone in different systems. The relevant parts of the respective plots are enlarged and arranged next to it in the same row with the proper scale.

A careful observation suggests that water has two sharp peaks (~ 0.2 and ~ 0.3nm) with 0.5nm of the protein backbone and their position remain almost the same for the four systems considered in our study. It is evident from Figure 4a that density of water around protein's backbone gets affected by the presence of urea and choline. For example, in PWU (Figure 4a, red curve), the presence of urea reduces the water density as the first peak height is obtained at ~ 0.1. However, in the PWC (Figure 4a, blue curve) system, a reverse trend is observed i.e. peak height increases to ~ 0.175 compared to PW system, where the g(r) is found to be ~ 0.125. It seems that bulky choline pushes the water molecule towards the protein backbone and thereby preserve the tightness of the protein structure. The preservation of the structure of a small peptide in the presence of another choline-containing osmolyte choline-o-sulphate is recently shown by Paul *et al.*[44] Considering the PWUC system (Figure 4a, green curve), water density greatly reduced surrounding the protein backbone. While probing the urea density around the protein backbone (Figure 4b), we find a sharp peak of urea molecules ~ 0.2nm, which signifies its direct interaction property of urea while denaturing the protein structure. The g(r) value of urea reduces from 0.15 to 0.1 while comparing PWUC (Figure 4b, green curve) system with PWU (Figure 4b, red curve). It proves that in the presence of bulky choline molecules, urea density is greatly reduced at the vicinity of the protein backbone. To account for the distribution of choline surrounding the protein backbone (Figure 4c), we do not find any kind of sharp peak with 0.5nm of the protein backbone, rather we get small hump. This can be attributed to the fact that choline does not have preferential interaction with the protein backbone.

### 4.2. Theory and calculation of Kirkwood-Buff integral (KBI)

Kirkwood and Buff developed a concrete molecular level theory regarding the behavior of solutions.[45] This theory mainly focuses on the detailed analysis of the accumulation of co-solute around a solute using the local/bulk partition model. Over the last few decades, the framework has been developed and implemented by several groups.[46–55] The spatial integral over the pair correlation function is designated as the Kirkwood-Buff integral (KBI) and has the following definition –

$$G_{ij} = 4\pi \int_0^\infty [g_{ij}(r) - 1] r^2 dr \tag{1}$$

where $g_{ij}(r)$ is the radial distribution function between the components i and j. KBI has been found to be extremely useful in studying the distribution of solvent or co-solvent around biomolecules.[56, 57] The plot of KBI for different substituents around the protein backbone is provided in the supplementary information (Figure S1). Difference between $G_{ij}$ and $G_{ik}$ i.e. ($G_{ij}$ - $G_{ik}$) is the preferential interaction of i with k over j. Values of ($G_{ij}$ - $G_{ik}$) of the protein and other solvent/co-solvent are depicted in Figure 5. If the value of ($G_{ij}$ - $G_{ik}$) is positive then the component i have a preference for component j over component k. Whereas, for a negative value, the preference is exactly the reverse.

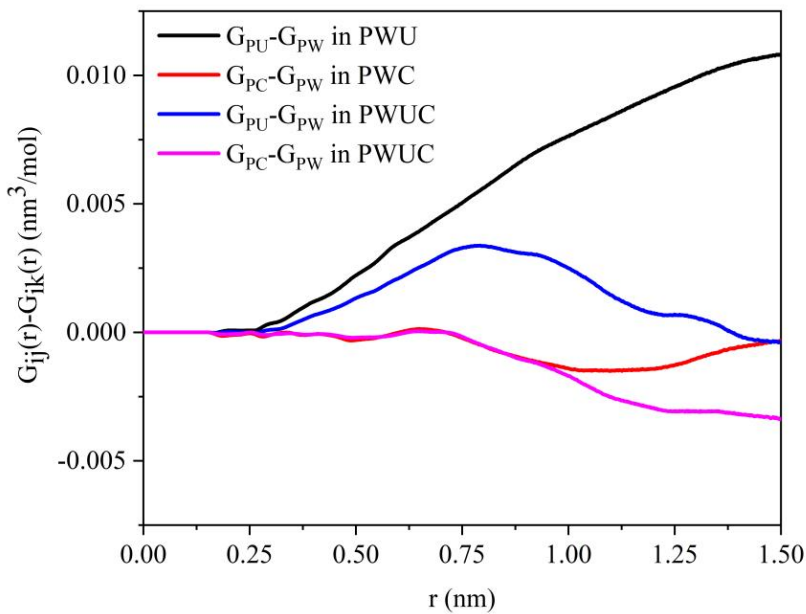

**Figure 5:** Preferential Kirkwood-Buff integrals ($G_{ij}(r)$-$G_{ik}(r)$) of co-solvents around HP-36 in different systems as a function of distance (r) from protein backbone, where i stands for protein backbone and j and k stand for the solvent/co-solvent considered as mentioned in the legend.

The preferential interaction of protein backbone with one co-solvent compared to the other can be estimated by computing the pair correlation functions of co-solvents around the protein backbone. Figure 5 clearly shows a positive value for $G_{PU}$ - $G_{PW}$ in PWU system (Figure 5, black curve). This signifies that the protein-urea interaction is manifold preferred over the protein-water interaction. Whereas in PWC (Figure 5, red curve), there is a net decrease in protein-choline interaction compared to protein-water interaction which actually indicates that choline influences the dynamics of water in such a way that water density around protein increases and choline gets excluded from the protein backbone. The preferential exclusion of the protecting osmolyte, namely TMAO, from the protein surface was reported earlier by different groups.[58, 59] This is also reflected in the radial distribution of water around protein's backbone in PWC system (Figure 4a, blue curve). In PWUC, the $G_{PU}$ – $G_{PW}$ (Figure 5, blue curve) value decreases compared to that in PWU (Figure 5, black curve), indicating the fact that in the ternary mixture, due to restriction of its movement urea cannot preferentially approach towards the protein surface. The $G_{PC}$ - $G_{PW}$ value in PWUC (Figure 5, magenta curve) shows a similar kind of trend as that in PWC up to 1nm distance and goes to a more negative value for larger r. This signifies that the interaction of choline with protein does not change much rather choline alters the preference of water and urea around the protein surface.

From the above results, it could be summarized that the preference of urea around protein is diminished in the presence of choline. Presence of choline actually promotes further accumulation of water molecules around the protein's surface, which helps in preserving the structure of the protein backbone. In a nutshell, the role of choline is two folds. On one hand, it excludes urea to approach protein by behaving as a nano-crowder. On the other hand, it helps in maintaining an adequate water density around the protein and thereby preserves its secondary structure.

## 5. Dynamics of solvent and co-solvent around protein

Analyses of KBI provide a static picture of the distribution of the solvent and co-solvent molecules around the protein. However, it does not provide information regarding the dynamics of these constituents, which in principle play a pivotal role in deciding the protein's conformation. The dynamic behavior of the solvent and cosolvent molecules around the protein is characterized by considering three important observables mean square deviation (MSD), rotational correlation and hydrogen bond autocorrelation function.

### 5.1. Translational motion of solvents and co-solvents

The translational motions of the constituent molecules can be measured from their mean square displacement (MSD) which is defined as,

$$\text{MSD} = \langle \Delta r(t)^2 \rangle = \langle (r(t) - r(0))^2 \rangle \tag{2}$$

where r(t) is the position of the center of mass of the molecule under consideration at time t. Only those molecules which are within a distance of 0.4 nm from the protein backbone were considered for MSD calculation. The results are depicted in Figure 6. For comparison, the MSDs for different systems are plotted along with the MSD of pure bulk water under identical condition (at temperature 310K and pressure 1atm). As expected, the presence of protein slows down the translational motion of water molecule in comparison to that in bulk. In addition, the presence of cosolvent further reduces the growth of MSD over time. This kind of restricted motions of water near the surface of biomolecules are well-known.[60-65] The translational motion of the water molecules in the presence of urea (red curve) is slower compared to the water molecules around the protein surface in the PW system (black curve). The mobility of water gets highly decreased while choline is present in the system (PWC, PWUC). We can see that the MSD curve water molecules near the protein surface for PWC (blue curve) and PWUC (black curve) nearly merges with the plotting x-axis, which ultimately indicates toward the extraordinary slow water molecules in those systems.

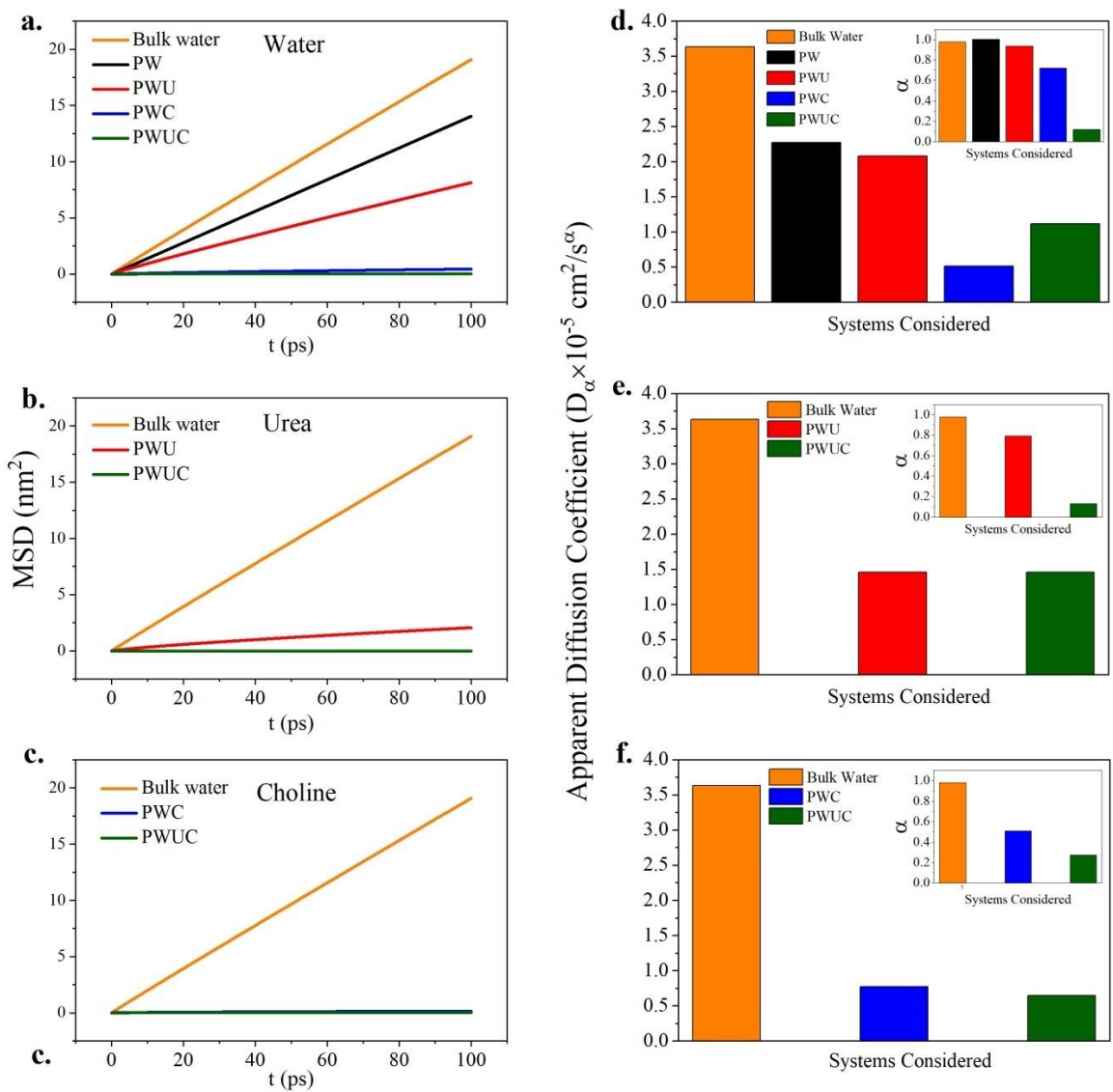

**Figure 6:** Mean Square Displacement (MSD) of a) Water, b) Urea, c) Choline for different systems considered in this investigation. For MSD of each constituent, the apparent Diffusion coefficient ($D_\alpha$) values are placed next to it.

We have also probed the translational motion of the co-solvents present in the system namely, urea and choline. We could see that due to the bulky size of the urea (Figure 6b) and choline (Figure 6c) their displacement is manifold slower than that of the water molecules, which is even prominent for the choline molecules (Figure 6c). Considering Figure 6b, urea molecule has some extent of translational motion in the PWU system (Figure 6b, red curve), which is diminished in the PWUC system (Figure 6b, green curve). Comparing the urea motion with that of the choline (Figure 6c) we find that choline has diminished motion even in the PWC system (Figure 6c, blue curve), which almost comparable with that of the PWUC system (Figure 6c, green curve).

To quantify the nature of the plots discussed above and to have the essence of the degree of slowness, we have also calculated the apparent diffusion coefficient ($D_\alpha$) from the MSD plots, where MSD scales as $t^\alpha$ –

$$\langle \Delta r^2 \rangle = 6 D_\alpha t^\alpha \qquad (3)$$

where $\alpha = 1$ represents the purely diffusive case and $\alpha < 1$ is termed as subdiffusion and $D_\alpha$ has a dimension $L^2/T^\alpha$. [64, 66, 67] True Diffusion coefficients (D) can be extracted from equation 3 only in the diffusive region (for $\alpha = 1$). The values of $D_\alpha$ and $\alpha$, extracted from the MSD plots are listed in Table 1.

**Table 1:** Values of apparent diffusion coefficients ($D_\alpha$) and the exponents ($\alpha$) for a) Water, b) Urea, c) choline in different systems considered in this investigation. The corresponding values for pure bulk water are given for comparison purpose.

| Water | | Parameters | |
|---|---|---|---|
| System | Starting Point (ns) | $D_\alpha (\times 10^{-5} cm^2/s^\alpha)$ | $\alpha$ |
| Bulk Water | - | 3.63 | 1.00 |
| PW | 190 | 2.27 | 0.98 |
| PWU | 190 | 2.08 | 0.94 |
| PWC | 190 | 0.51 | 0.72 |
| PWUC | 490 | 1.11 | 0.12 |
| Urea | | Parameters | |
| System | Starting Point (ns) | $D_\alpha (\times 10^{-5} cm^2/s^\alpha)$ | $\alpha$ |
| PWU | 190 | 1.46 | 0.79 |
| PWUC | 490 | 0.73 | 0.13 |
| Choline | | Parameters | |
| System | Starting Point (ns) | $D_\alpha (\times 10^{-5} cm^2/s^\alpha)$ | $\alpha$ |
| PWC | 190 | 0.77 | 0.51 |
| PWUC | 490 | 0.65 | 0.27 |

The value of $\alpha$ is found to be very close to 1 for pure bulk water which signifies diffusive behavior. The MSD for bulk water (orange) is plotted along with MSDs of other constituents for comparison. In the presence of the protein, the translational motion of water slows down as reflected in the small deviation of $\alpha$ from 1. This nearly diffusive behavior of water around the protein is also evident from other studies.[68]
The addition of cosolvents further reduces the translational motion of water and the degree of slowness depends on the nature of the cosolvent. Presence of urea makes the water around protein weakly subdiffusive in nature. Introduction of bulky choline ion induces further restriction in waters motion and the MSD plot shows moderately subdiffusive behavior. Water molecules around protein have least translational movement in the ternary mixture and behave to be strongly subdiffusive (Figure 6a).

Due to strong interaction with proteins backbone urea molecules show moderately subdiffusive behavior in PWU system. The restriction in the movement of components of the ternary mixture makes urea to be strongly subdiffusive in PWUC (Figure 6b). The choline ion in both PWC and PWUC systems shows strongly subdiffusive nature of MSD owing to its slower translational motion (Figure 6c).

## 5.2. Calculation of rotational autocorrelation function

To elucidate the internal dynamics of different constituents in a mixture, rotational motion is considered as one of the important measures. In this section, we focus on the rotational motion of the water, urea, and choline molecules. In order to quantify the rotational motion, we compute the following rotational autocorrelation function –

$$C_\mu(t) = \frac{\langle \hat{\mu}(0)\hat{\mu}(t) \rangle}{\langle \hat{\mu}(0)^2 \rangle} \qquad (4)$$

where $\hat{\mu}(t)$ indicates the resultant dipole of the molecule at time t. The vectors along resultant dipole for the three moieties are shown in Figure 7. Physically the above autocorrelation function describes how dipole-dipole correlation decays over time due to rotation. Thus the decay time corresponds to the orientation time of the molecules concerned. We use the following bi-exponential function to fit the data –

$$C_\mu(t) = A_1 e^{-t/\tau_1} + A_2 e^{-t/\tau_2} \qquad (5)$$

where $\tau_1$ and $\tau_2$ represent the rotational relaxation times for the initial faster and subsequent slower decay respectively. $A_1$ and $A_2$ represent the contribution of the corresponding timescales. The values are listed in table 2. The rotational decays for the three constituents in different systems are plotted with that of bulk water for comparison purpose (Figure 7, orange curve). A careful observation of the values in Table 2 reveals that the slower relaxation time ($\tau_1$) contributes more in describing the overall nature of the decay curve. Therefore we consider the values of $\tau_2$ for comparing the relative rotational motion of the moieties.

As expected, Bulk water molecules show the fastest rotational decay with a timescale around 4.38ps. [61] The water molecules around the protein in PW and PWU systems have ~3 times slower rotational relaxation compared to those in bulk water because of the water-protein interaction. In the presence of choline, a layer of water is accumulated around protein surface which leads to slowing down of their rotation, which has been reflected in the higher value of $\tau_2$ (~25 times higher than $\tau_2$ in bulk water) (Fig 7a). The spatial orientation of the three components in the ternary mixture creates a crowded layer around the protein. In that environment rotational relaxation of water appears to be extremely slow as reflected in the high value of $\tau_2$. The restricted rotational dynamics of water around protein surface was evident from previous experimental and computational studies.[69]

Due to substantial interaction of urea with protein, its orientational dynamics also gets affected and it follows a decay slower than bulk water as well as the water around the protein (Figure 7b). $\tau_2$ is 7 times higher than bulk water. Due to bulkier size, choline molecules around protein take a sufficiently long time to change its orientation and the rotational dynamics time scale becomes sufficiently large ($\tau_2 \approx 80$ps) (Figure 7c).

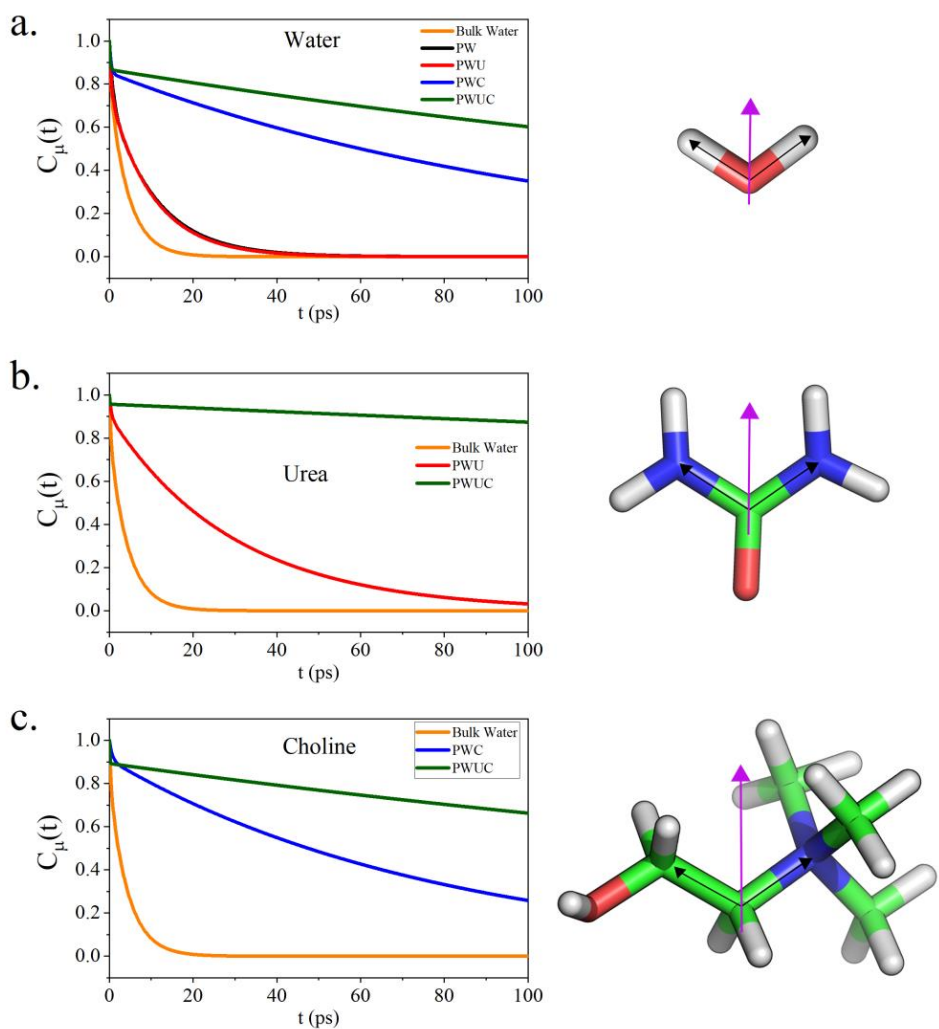

**Figure 7:** Rotational auto-correlation function, $C_\mu(t)$ of a) Water, b) Urea, c) Choline around the protein backbone for different systems considered in this investigation. At the right-hand side panel plots, respective molecules for which $C_\mu(t)$ is calculated are provided. The resultant dipole with respect to which $C_\mu(t)$ is calculated is presented in magenta. Considering the molecules, Carbon is represented in Green, Nitrogen in Blue, Oxygen in Red, Hydrogen in White.

Presence of water, urea, and choline chloride in the PWUC slow down the dynamics of all the components forming a tri-component crowded layer around the protein. This leads to extremely slow orientational decay of urea and choline as reflected in very large values of $\tau_2$ in PWUC (in the range of 1000 ps and 330 ps respectively) (Figure 7b and 7c).

**Table 2:** Relaxation time of the rotational auto-correlation ($C_\mu(t)$) for the constituent a) Water, b) Urea, c) Choline around the protein's backbone.

| Constituent | System | Relaxation Time ($\tau$) (ps) | Contribution (%) |
|---|---|---|---|
| a. Water | Bulk Water | 0.38 | 18 |
| | | 4.38 | 82 |
| | PW | 0.85 | 26 |
| | | 11.04 | 74 |
| | PWU | 0.51 | 24 |
| | | 10.44 | 76 |
| | PWC | 0.37 | 15 |
| | | 112.68 | 85 |
| | PWUC | 0.19 | 13 |
| | | 274.39 | 87 |
| b. Urea | PWU | 0.44 | 10 |
| | | 29.79 | 90 |
| | PWUC | 0.08 | 4 |
| | | 1099.69 | 96 |
| c. Choline | PWC | 0.69 | 9 |
| | | 79.38 | 91 |
| | PWUC | 0.11 | 11 |
| | | 335.51 | 89 |

## 5.3. Hydrogen bond (HB) autocorrelation function

Since the two co-solvents considered here (urea and choline) have polar functional groups, their interactions with the protein are guided by hydrogen bonds with the surface residues of the protein. To understand the change in hydrogen bond property between HP-36 and water/urea/choline for different systems considered here, the hydrogen bond (HB) dynamics are monitored by computing the hydrogen bond auto-correlation function, $C_{HB}(t)$.[12,70,71] A pair of acceptor and donor is considered to form hydrogen bond if the distance between the donor and acceptor atom is less than 0.35nm and the acceptor-donor-hydrogen angle is less than 30°.[32] The auto-correlation function is computed according to the following formula[12,70]-

$$C_{HB}(t) = \frac{\langle h(0)h(t) \rangle}{\langle h(0)^2 \rangle} \qquad (6)$$

Where, h(t) is the hydrogen bond lifetime function, which is considered to be 1 if a pair of donor-acceptor is forming a hydrogen bond at time t and 0 if it is not formed at time t. The auto-correlation function is then fit into a bi-exponential decay function in consistence with previous studies on water dynamics[72]-

$$C_{HB}(t) = B_1 e^{-t/\tau_1^{HB}} + B_2 e^{-t/\tau_2^{HB}} \qquad (7)$$

Here, $\tau_1^{HB}$ and $\tau_2^{HB}$ represent the hydrogen bond relaxation times, whereas $B_1$ and $B_2$ represent the contribution of the corresponding timescales. Since the interest of the present study is the dynamics of solvent and co-solvent in the context of protein's conformational change only those constituents (water/urea/choline) which are close to protein backbone (at a distance 0.4nm from the protein backbone) are considered for calculation of $C_{HB}(t)$.

From the bi-exponential fitting of $C_{HB}(t)$, we get two relaxation times (i) $\tau_1^{HB}$ which describes the initial sharp fall in the plot and (ii) $\tau_2^{HB}$ which corresponds to the relatively slow decay after the initial drop in $C_{HB}(t)$. In literature, the shorter time scale $\tau_1^{HB}$ is attributed to libration of water molecules and inter-oxygen vibration.[12, 73]

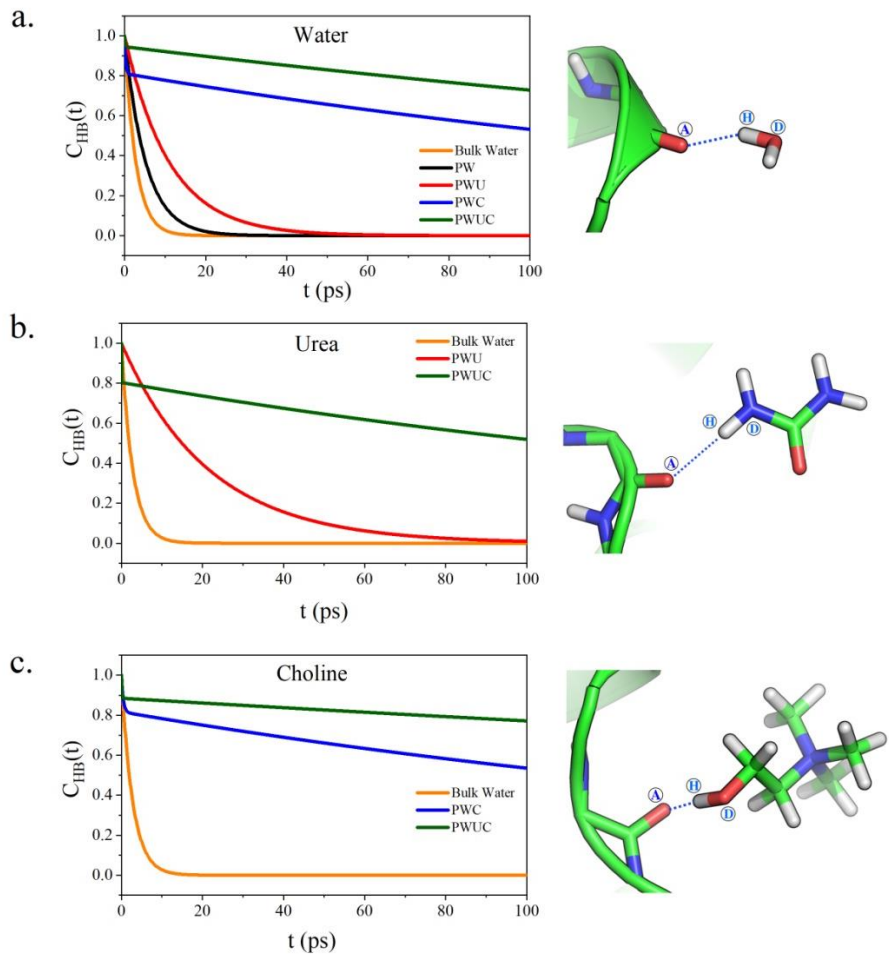

**Figure 8:** Hydrogen bond auto-correlation function $C_{HB}(t)$ of a) Water, b) Urea, c) Choline around the protein backbone for different systems considered in this investigation. At the right-hand side panel, respective molecules for which $C_{HB}(t)$ is calculated is provided. A=Acceptor, H=Hydrogen, D=Donor. Hydrogen bonds are shown in dotted line.

However, the other time scale $\tau_2^{HB}$ which is longer, corresponds to hydrogen bond relaxation time.[12, 73] Therefore we consider $\tau_2^{HB}$ as the most important time scale to describe the water dynamics and compare its value for different systems. It is evident from Figure 8 that the value of $\tau_2^{HB}$ of water around protein backbone (Figure 8a, black curve) is higher compared to the bulk water (Figure 8a, orange curve) because of relatively stronger hydrogen bond interaction between structured protein backbone and water molecules within the first solvation shell. Table 3 shows that the obtained $\tau_2^{HB}$ value of water is 5 times slower in the presence of protein than that of the bulk water. While considering the PWU system (Figure 8a, red curve), we find that the hydrogen bond formation and breaking kinetics is even slower in the presence of urea, which is clear from the higher relaxation time (both $\tau_1^{HB}$ and $\tau_2^{HB}$) of water compared to that of the PW system. This is consistent with the fact that as the protein unfolds gradually upon addition of urea, the water molecules interact strongly with the exposed backbone atoms and relax slowly (higher $\tau_2^{HB}$ value). Bandyopadhyay *et al.* have shown that urea influences the dynamics of water more but does not affect the water structure.[74] The dependence of $C_{HB}(t)$ on the conformation of protein was also reported by Rani *et al.* around intrinsically disordered protein with extended conformation.[72]

**Table 3:** Time constant of the hydrogen bond auto-correlation $C_{HB}(t)$ for the hydrogen bond between protein's backbone and a) Water, b) Urea, c) Choline

| Constituent | System | Relaxation Time (τ) (ps) | Contribution (%) |
|---|---|---|---|
| a. Water | Bulk Water | 0.32 | 23 |
| | | 3.53 | 77 |
| | PW | 0.29 | 75 |
| | | 19.96 | 25 |
| | PWU | 1.48 | 49 |
| | | 20.48 | 51 |
| | PWC | 0.26 | 19 |
| | | 237.06 | 81 |
| | PWUC | 0.16 | 6 |
| | | 383.32 | 94 |
| b. Urea | PWU | 0.36 | 40 |
| | | 35.96 | 60 |
| | PWUC | 0.04 | 20 |
| | | 228.89 | 80 |
| c. Choline | PWC | 0.45 | 18 |
| | | 236.53 | 82 |
| | PWUC | 0.11 | 12 |
| | | 729.82 | 88 |

In this case, they showed that the unstructured protein slows down the HB dynamics between water and the protein's backbone. The dynamical heterogeneity in the dynamics of water molecule around the protein and that in bulk were demonstrated by Arya *et al.* using both experimental and computational approach.[75] The role of urea to slow down the water dynamics has been recently described by Ojha *et al.* also.[76] Addition of bulky choline ion affects the HB dynamics of water around protein backbone (both in PWC and PWUC) by manifold and hence it is reflected clearly in the timescale. In PWC (Figure 8a, blue curve), choline affects the dynamics of water in such a way that it facilitates the accumulation of the water molecules on the protein surface, which is evident from the g(r) plots (Figure 4a). This results in the confinement of the hydration layer water molecules, where hydrogen bond relaxation is around 70 times slower than that of the bulk water. In the ternary mixture (Figure 8a, green curve) due to the presence of two co-solvents, water molecules are within a sterically restricted environment which slows down the dynamics in general (~ 100 times). This kind of slowly moving water molecules and their subsequent confined motion is also evident from the ultra-slow translational motion of water (Figure 6a).

The dynamics of the co-solvents (urea and choline) is also monitored using $C_{HB}(t)$. For urea, hydrogen bond is considered between the protein backbone oxygen (>C=O) and the amine hydrogen of urea (-NH$_2$) and for protein-choline HB, the hydrogen bond between backbone oxygen (>C=O) of protein and the hydroxyl group (-OH) of choline is considered. In PWU (Figure 8b, red curve), the observed urea relaxation timescale is around 10 times slower than that of the bulk water, which further slows down to ~229ps in the ternary mixture (Figure 8b, green curve). This attributes to the fact that the addition of bulky choline to urea (PWUC system) slows down the dynamics of urea by almost 6 times. Figure 8c represents the HB dynamics of choline molecule within the first solvation shell of protein. It seems that the bulky choline molecules hinder the movement of other constituents present with it (Figure 8a) and also its own motion is also highly restricted. As a result, the HB relaxation timescale of choline becomes 67 times and 200 times slower in PWC (Figure 8c, blue curve) and PWUC (Figure 8c, green curve) system respectively.

## 6. Effect of choline chloride (ChCl) imposed crowding on the protein molecule

### 6.1. Root mean square fluctuation (RMSF)

To investigate the effect of choline molecules on the protein structure, we compare the residue level fluctuations by calculating the root mean square fluctuation (RMSF) of each residue. RMSF is the time-averaged root mean square deviation (RMSD) of an entity and is calculated as,

$$\text{RMSF} = \sqrt{\left[\frac{1}{t_f}\sum_{t=0}^{t_f}(r(t)-r(0))^2\right]} \tag{8}$$

where r(t) is the position of the alpha carbon (Cα) of an amino acid residue at time t.

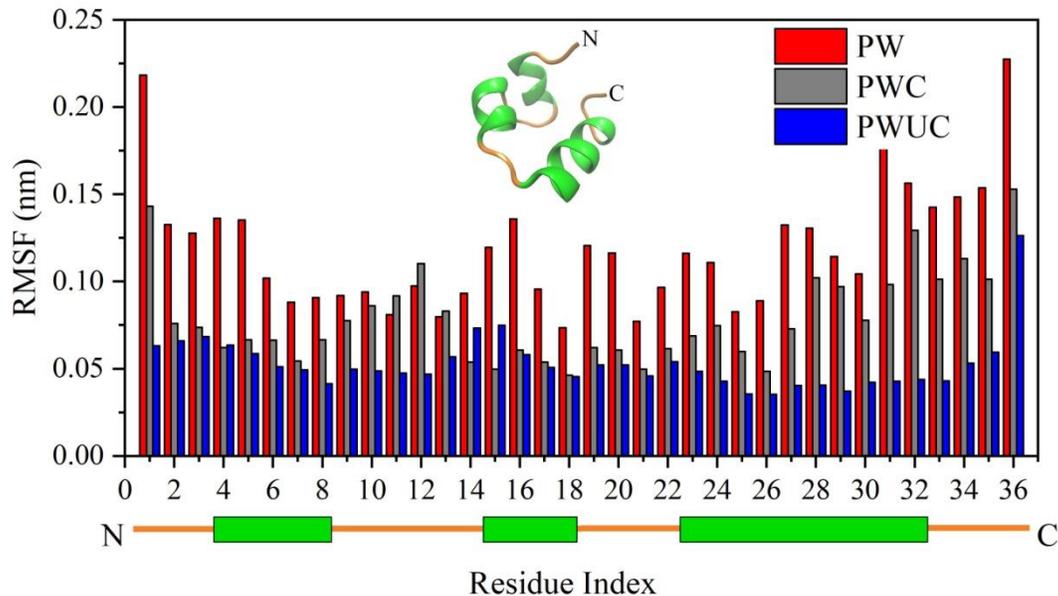

**Figure 9:** Root mean square fluctuation (RMSF) of the residues of HP-36 for PW, PWC, PWUC presented in a bar diagram in red, grey and blue color respectively. The helices and the coil in the 3D representation of the small protein are colored in green and orange respectively.

RMSF of the three systems, PW, PWC, and PWUC were plotted together in Figure 9 to investigate and compare the structural fluctuation of the amino acid residues. It is clear from Figure 9 that for all the systems the terminal residues have the maximum fluctuations which are quite obvious because of the apparent free movement of those residues. The values also suggest that the addition of choline in the system makes the protein structurally rigid, which is reflected in the lower RMSF values of the amino acid residues in the PWC and PWUC systems. In the presence of ionic liquids, other biomolecules like DNA has also been found to adopt this kind of rigidity.[77] The Bulky choline molecules introduce crowding to the systems, which was not present in the PW system. An average fluctuation of 0.1nm is observed for the PW system, while it is around 0.08nm for PWC and 0.04nm for the PWUC system. The probable reason for the lowest observed RMSF in the PWUC system is because of the high viscosity in the medium exerted by the slow coupled movement of the three constituents.

### 6.2. Conformational flexibility of the Protein structure

To monitor the conformational flexibility of HP-36, two important structural parameters, namely, root mean square deviation (RMSD) and radius of gyration ($R_g$) are chosen, which are capable of describing the conformational space effectively. We have constructed a free energy surface map from the density of the distribution of the aforesaid structure parameters. The probability of occurrence of the points within 0.1 nm × 0.1 nm grid is calculated from the RMSD vs $R_g$ plots as follows –

$$P_{ij} = \frac{n_{ij}}{N_{ij}} \tag{9}$$

Where $n_{ij}$ is the Number of conformations within an area confined between RMSD values i and i+Δi and $R_g$ values j and j+Δj. $N_{ij}$ is the total number of conformations. Free energy change is then calculated using the formula –

$$\Delta G_{ij} = -RT \ln P_{ij} \tag{10}$$

The free energy surface is presented in Figure 10 (a-d) for PW, PWU, PWC, PWUC systems respectively where an increase in free energy change is depicted using a color gradient (from blue to cyan). The blue-colored region describes the most populated region in the free energy surface. For the PW system, the conformations are distributed in the region having RMSD below 0.4 nm and $R_g$ below 1.0 nm.

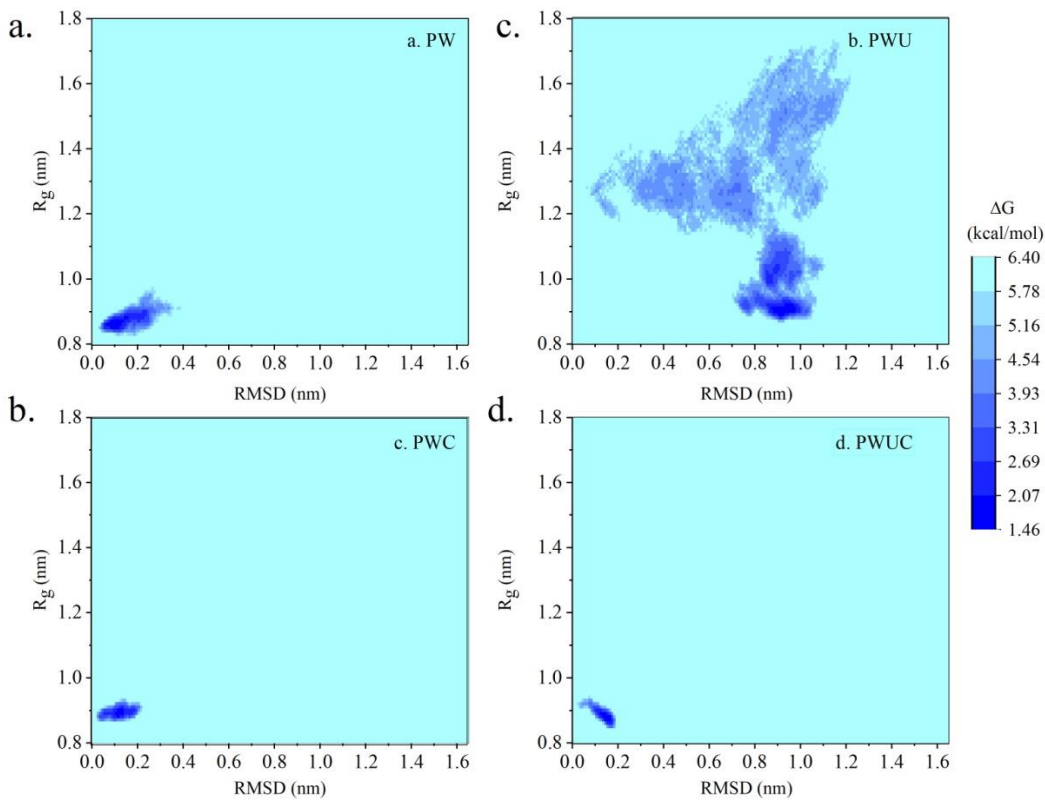

**Figure 10:** Contour map of the free energy profile obtained from the RMSD and $R_g$ distribution of (a) PW, (b) PWU, (c) PWC, (d) PWUC, considering the $C_\alpha$ of protein's backbone throughout the simulation.

It is evident from Figure 10b that in presence of urea, it undergoes a huge structural change starting from the fully folded ensemble to a coil-like structure. As the structural parameters are concerned, conformations of protein are sampled over a wide range in the RMSD-$R_g$ space. This also shows that the most stable conformation of the protein molecule corresponds roughly to RMSD value of ~0.9-1.0nm and $R_g$ value of~0.5nm. The most probable structure corresponds to an RMSD value ~0.1nm and $R_g$ value ~0.3nm. Introduction of choline molecules in the system (for both PWC and PWUC represented in Figure 10c and Figure 10d respectively) impose a crowded environment surrounding the protein that ultimately leads to enhancement in the structural integrity of the protein molecule and creates hindrance in its free movement. Figure 10c and Figure 10d clearly indicates, while comparing the protein structure even with the PW system (represented in Figure 10a), it seems to be much more rigid and sampled over very little RMSD-$R_g$ space.

## 7. Conclusions

In the present investigation, using extensive MD simulations we find how a 36-residue protein (HP-36) maintains reasonable conformational stability in the ternary mixture of water, urea, and choline chloride (ChCl) although urea is known to denature the protein structure. The ammonium-based stabilizer choline chloride has been found to shield the secondary structure of the protein from the action of the denaturant urea at room temperature. A critical assessment of change of the secondary structure in different environments exhibits the preservation of protein structure in the deep eutectic condition whereas sufficient secondary structure loss was observed in presence of urea (Figure 2 and 3).In order to gain molecular-level insights into the counteraction of the urea-induced denaturation in the ternary mixture, we have calculated g(r) of the water, urea, and choline surrounding the protein backbone (Figure 4). The preferential occupancy of the water, urea, and choline is monitored by means of Kirkwood-Buff integrals (Figure 5). Calculation of these distribution functions indicates that the preferential accumulation of urea over water around protein backbone is sufficiently reduced in the presence of choline chloride. Additionally, choline chloride modulates the water molecules in such a way that density of water increases around the protein backbone compare to that in other systems which in turn helps to retain its native structural fold.

In order to delineate the effect of the dynamics of the solvent and co-solvent towards determining the solvation property of the system, three important observables are calculated *viz*. mean square deviation (measure of translational dynamics), orientational correlation (measure of rotational dynamics), and hydrogen bond autocorrelation (a measure of dynamics of protein-solvent/cosolvent interaction). A comparison of these three quantities for all the components of the ternary mixture reveals interesting features. Water, urea, and choline chloride, all three of them have subdiffusive MSD where the α values vary between 0.15 and 0.3 (Figure 6). Although the decay in the rotational correlation of water, urea, and choline chloride in different systems (Bulk, PW, PWU, and PWC) have varying time scales, all of them exhibit extremely slow rotation in the PWUC system ($\tau_2$ is more than 270 ps) (Figure 7). Similar behavior is observed for HB autocorrelation. The relaxation time of the hydrogen bonds between protein backbone and the respective polar hydrogen of the solvent/cosolvent molecule ($\tau_2^{HB}$) is quite long for all the three components and varies in a similar range (Figure 8). The sluggish translation, rotational and hydrogen bond dynamics for the here components in PWUC system confirms the correlated slowing down of all the three components around the protein. This correlated restriction of molecular motion can also be attributed to the spatial arrangement of the bulky molecules of denaturant and the protecting osmolyte. This kind of slower motion of water and other co-solvents around a protein is also reported in earlier for different denaturants or protecting osmolytes.[66, 78] This type of dynamics modulates the rigidity of the surrounding environment and turns it into a slowly moving solvent shell around the protein. As a result a crowding effect is imposed on the protein to reduce its conformational flexibility in the ternary mixture and even in PWC system also reflected in the RMSD-vs-$R_g$ plot (Figure 10).

The findings of this investigation summarize that the role of the bulky choline ion can be viewed as that of a nano-crowder [79], which suppress the dynamics of the protein and the other co-solvents, thereby preventing the unfolding. The role of choline can be compared to the role of another osmolyte TMAO, which also acts as a nano-crowder and results in entropic stabilization of the protein.[10] We hope that the current work is extremely useful in understanding the role of protecting osmolytes at the microscopic level and will contribute in designing potential candidates for this purpose.

## Supplementary Information

A brief overview of the simulated systems, Calculation of Kirkwood-Buff integral (KBI) of different constituent around protein backbone, Time dependence of KBI.

## Author Information

### Corresponding Author

* Email: rajarshi@chem.iitb.ac.in
Mailing Address: Department of Chemistry, Indian Institute of Technology Bombay, Powai, Mumbai-400076, India.
Phone: + 91-022-2576 7192.

Fax: + 91-022-2576 7152.

## Disclosure Statement

The authors declare no competing financial interest.

## Acknowledgment

A.M. and S.S. thanks IIT Bombay and L.T. thanks University Grant Commission, Govt. of India for fellowship. The authors acknowledge the computational facility (HPC) provided by the institute.

# Choline chloride as a nano-crowder protects HP-36 from urea-induced denaturation: Insights from Solvent Dynamics and Protein-Solvent interaction

Atanu Maity, Soham Sarkar, Ligesh Theeyancheri, and Rajarshi Chakrabarti[*]

Department of Chemistry, Indian Institute of Technology Bombay, Powai, Mumbai-400076

E-mail: rajarshi@chem.iitb.ac.in

## 1. System information

| System Identifier | No. of water molecules | No. of urea molecules | No. of choline ions | No. of chloride ions | Simulation length (ns) |
|---|---|---|---|---|---|
| PW | 7974 | 0 | 0 | 2 | 200 |
| PWU | 4193 | 1180 | 0 | 2 | 200 |
| PWC | 3332 | 0 | 590 | 592 | 200 |
| PWUC | 851 | 1180 | 590 | 592 | 500 |

**Table S1:** Details of the composition of solvent/co-solvent in different systems and simulation Length

## 2. Calculation of Kirkwood-Buff integral (KBI)

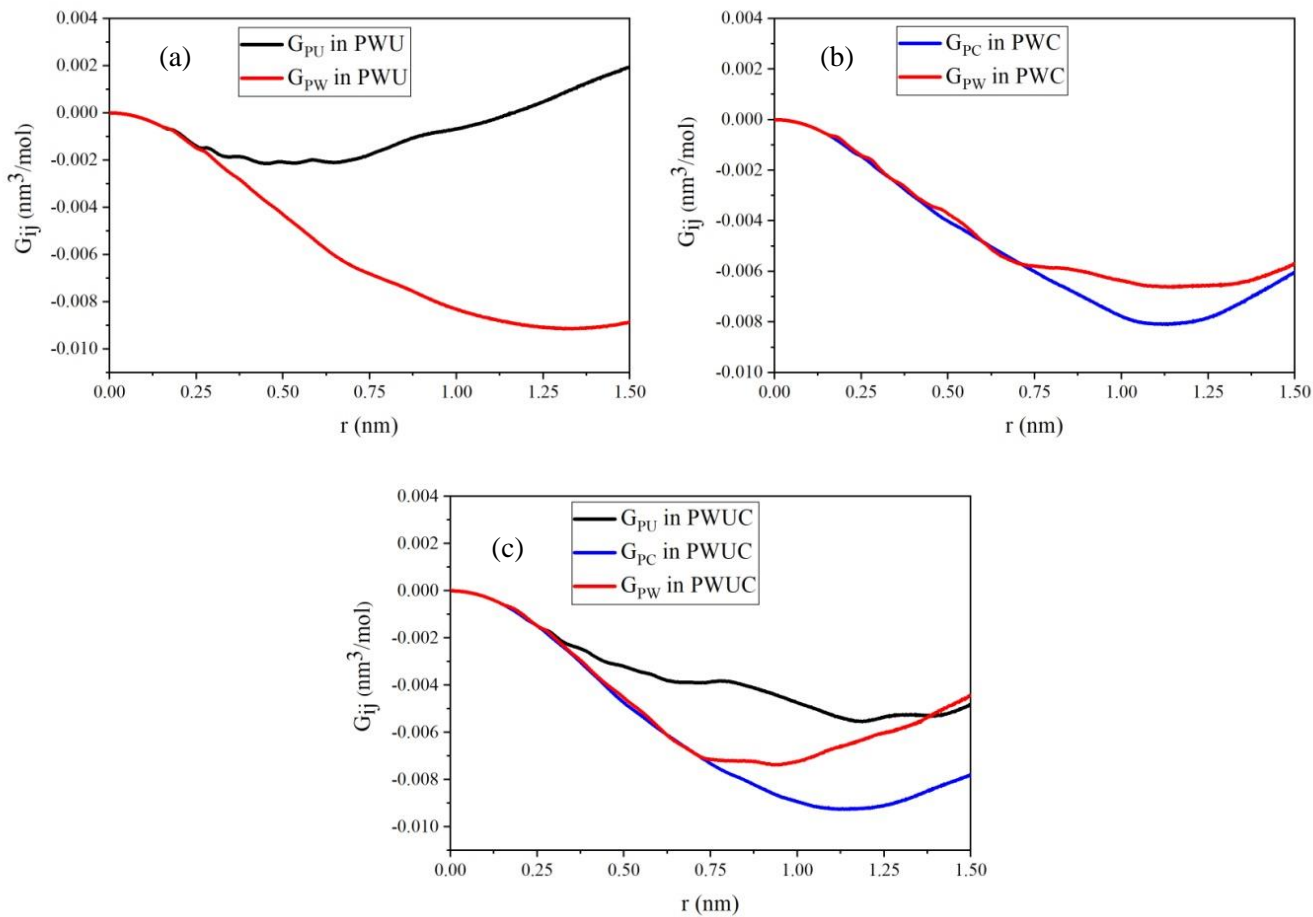

**Figure S1:** Kirkwood-Buff integrals ($G_{ij}(r)$) of water/urea/choline around protein for (a) PWU, (b) PWC, (c) PWUC as a function of distance (r) from protein backbone, where i stands for protein backbone and j stand for water/urea/choline considered and as mentioned in the legend. For different system considered here, KBI of water, urea, and choline are represented in red, black and blue color respectively.

## 3. Time dependence of Kirkwood-Buff integral (KBI)

The change in the preferential KBI with simulation time can be monitored by computing its value at an interval of 25ns for PW, PWU, PWC systems and at an interval of 50ns for PWUC system.

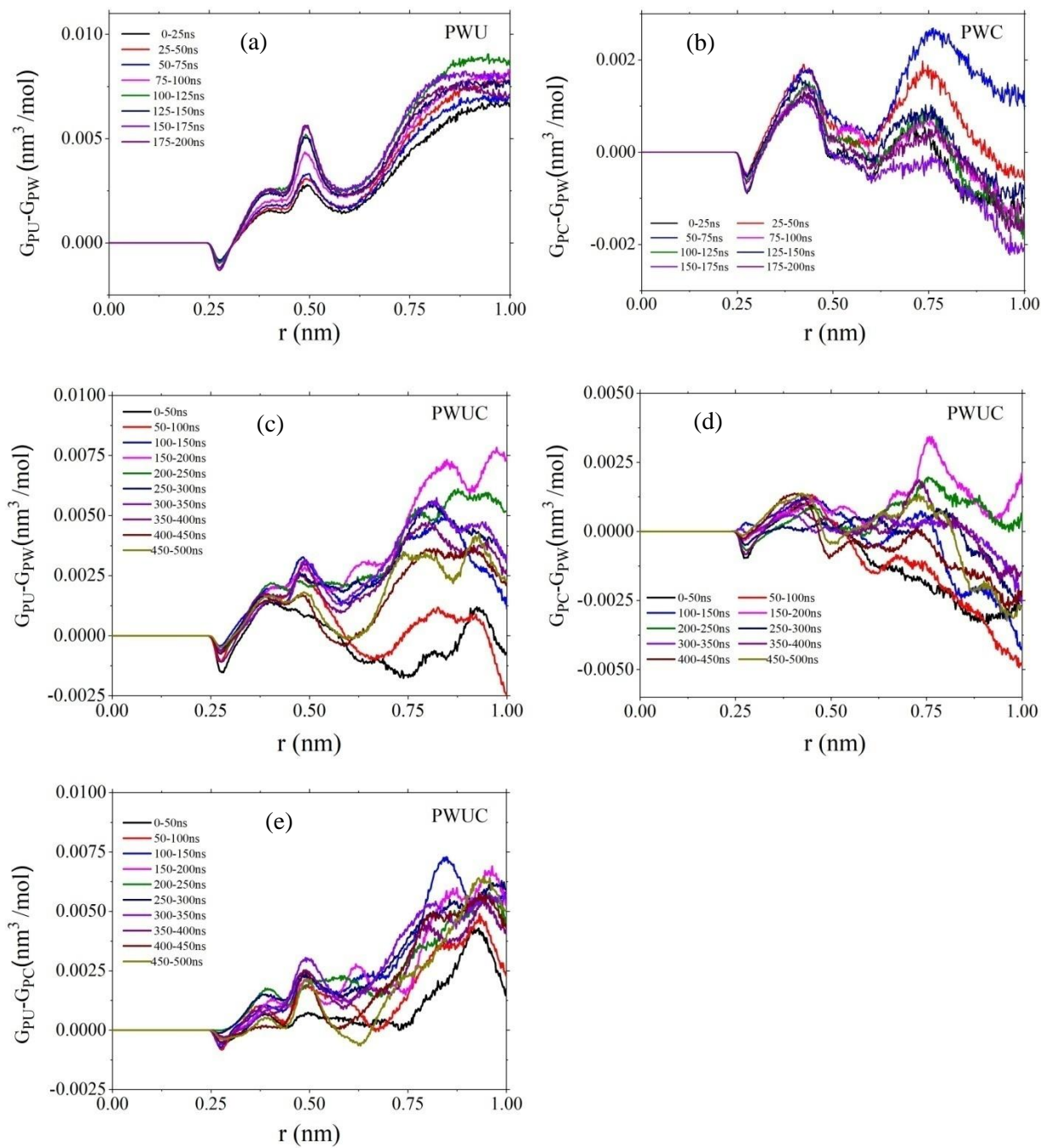

**Figure S2:** Change in Preferential KBI with simulation time considered for different systems.

Considering the time dependence of preferential KBI for PWU system (FigureS2, a) the value of $G_{PU} - G_{PW}$ increases with as simulation proceeds, which is clearly evident from the rising peak near 0.5nm. The time evolution of $G_{PC} - G_{PW}$ (FigureS2, b) reveals that the value decreases gradually as the first peak around 0.4nm decreases as time passes and it tends to go to negative for larger distance from the protein backbone. This is a reflection of the gradual removal of choline from the protein backbone and an increase in water in the hydration layer. Considering the PWUC system (FigureS2, c, d, e), the sharp rise of $1^{st}$ peak of urea density surrounding the protein (FigureS2, a) is get diminished in the presence of choline molecules (FigureS2, c). Considering other combinations in the PWUC system, we find the same kind of trend of choline for the protein backbone over water molecules (FigureS2, d) as we have found earlier (FigureS2, d). Beside that we have calculated the preferential KBI of urea over choline for protein backbone. We find it positive, which signifies a higher affinity of urea towards protein backbone compares to choline.